\documentclass[aps,prd,floatfix,nofootinbib]{revtex4} 
\usepackage{graphics}
\usepackage{graphicx}

\usepackage{latexsym}
\usepackage[cp866]{inputenc}
\usepackage[english]{babel}
\input epsf

\begin{document}

\title{\boldmath Phenomenological description of the $D^*_{s0}
(2317)\to D_s\pi^0$ decay}
\author{N. N. Achasov\,\footnote{achasov@math.nsc.ru}
and G. N. Shestakov\,\footnote{shestako@math.nsc.ru}}
\affiliation{\vspace{0.2cm} Laboratory of Theoretical Physics, S. L.
Sobolev Institute for Mathematics, 630090, Novosibirsk, Russia}

\begin{abstract}
For coupled channels $D^0K^+$, $D^+K^0$, $D_s^+\eta$, and
$D_s^+\pi^0$, the $S$-wave scattering amplitudes are constructed
taking into account the mixing of the isoscalar resonance
$D^*_{s0}(2317)^+$ with nonresonance amplitudes with isospin $I=1$.
The phenomenological approach we use allows us to quite simply clear
up  the general structure of the $D^*_{s0}(2317)^+\to D_s^+\pi^0$
decay  amplitude violating isospin. We show that the phase of this
amplitude coincides with the phase of the nonresonanct $D_s^+\pi^0 $
scattering amplitude in agreement with the Watson theorem. Its
modulus squared, as it should be, determines the width of the
resonance peak in the $D_s^+\pi^0$ channel. Taking into account the
$\pi^0-\eta$ mixing in internal lines up to the second order
inclusively ensures that the unitarity condition is fulfilled. The
presented analysis complements the description of the $D^*_{s0}(2317
)^+\to D_s^+\pi^0$ decay based on the coupled channel unitarized
chiral perturbation theory. The numerical estimates obtained by us
for the $D^*_{s0}(2317)^+\to D_s^+\pi^0$ decay width do not
contradict those available in the literature.
\end{abstract}

\maketitle

\section{Introduction}

The isospin violating decay of the $D^*_{s0}(2317)^\pm$ state with
$I(J^P)=0(0^+)$ into $D_s^\pm\pi^0$ \cite{Au03,Au06,Ab18,PDG24} has
been a unique ground for fruitful theoretical studies for over 20
years.\cite{Go03,CF03,Fa07,LS08,Gu08,Li13,Cl14,Gu18,Fu22,
Gu24,Yu25}. The main sources of isotopic invariance violation in
$D^*_{s0}(2317)^\pm\to D_s^\pm\pi^0$ decays are the mass differences
between charged and neutral $D$-mesons and kaons, and the
$\pi^0-\eta $ mixing \cite{Go03,CF03,Fa07,LS08,Gu08,Li13,
Cl14,Gu18,Fu22,Gu24,Yu25}. The most popular scheme for calculating
the amplitudes of the $D^*_{s0}(2317)^+$ production with its
subsequent decay into $D_s^+\pi^0$ is the coupled channel unitarized
chiral perturbation theory adapted to describe the interactions of
the light pseudoscalar mesons with the open charm ones, see for
review \cite{Gu08,Li13,Cl14,Gu18,Fu22,OM01,Gu06,Ga07,Gu09,Os24,Gu24}
and references therein. The discovery of $D^*_{s0}(2317)^+$ was
great luck for this scheme. The point is that the unknown
subtraction constant present in it can be selected so that in the
unitarized amplitudes under the thresholds of the coupled $DK$ and
$D_s\eta$ channels there appears a pole corresponding to the bound
state with the mass of the experimentally observed
$D^*_{s0}(2317)^+$ phenomenon \cite{Gu08,Li13,Cl14,Gu18,Fu22,OM01,
Gu06,Ga07,Gu09,Os24,Gu24}. Such a dynamically generated state in the
system of $DK$ and $D_s\eta$ pairs is often referred to as a
hadronic molecule \cite{Gu06, Ga07,Gu09,Gu24,Os24}. According to one
of the latest theoretical estimates \cite{Fu22}, the decay width
$\Gamma(D^*_{s0}(2317 )^+\to D_s^+\pi^0)$ is $(132\pm7)$ keV of
which the contribution from the $\pi^0-\eta $ mixing accounts for
$(20\pm2)$ keV, the contribution from the difference between the
$D^0K^+$ and $D^+K^0$ loops accounts for $(50\pm3)$ keV, and the
rest is due to constructive interference between the two isospin
violation mechanisms. Although the estimates of individual
contributions to $\Gamma(D^*_{s0} (2317)^+\to D_s^+\pi^0)$
\cite{Go03,CF03,Fa07, LS08,Gu08,Li13, Cl14,Gu18,Fu22,Gu24,Yu25} show
a noticeable scatter, they agree with each other in order of
magnitude. From the experiments it is known that the total width of
the $D^*_{s0}(2317)^+$ phenomenon $\Gamma<3.8$ MeV and its decay
fraction to $D_s^+\pi^0$ is $(100^{+ \,\,0}_{-20})$\% \cite{PDG24}.

In this paper, we use a phenomenological approach to constructing
the amplitudes of processes involving the $D^*_{s0}(2317)^+$
phenomenon which has isospin $I=0$ \cite{PDG24} and at the same time
the only kinematically admissible hadronic decay to the channel
$D_s^+\pi^0$ with $I=1$ violating isotopic invariance. The presence
of the $c\bar s$ pair in the quark structure of the $D^*_{s0}(2317
)^+$ state indicates its possible significant couplings with the
closed $D^0K^+$, $D^+K^0$, and $D_s^+\eta$ channels, as well as, due
to the $\pi^0-\eta$ mixing, with the decay channel into
$D_s^+\pi^0$. The manifestation of the $D^*_{s0}(2317)^+$ in the
amplitudes of the indicated channels is described by simple
resonance type expressions in Sec. II. In this Section, we also
construct simple expressions for the $S$-wave nonresonance
scattering amplitudes with $I=1$ associated with the $D^0K^+$,
$D^+K^0$, and $D_s^+\pi^0$ channels, as well as, due to the
$\pi^0-\eta$ mixing, with the $D_s^+\eta$ channel. The scattering
amplitudes taking into account the mixing of the isoscalar resonance
$D^*_{s0}(2317)^+$ with nonresonance amplitudes with isospin $I=1$,
are constructed in Sec. III. The obtained formulas for the complex
of mixed amplitudes with $I=0$ and 1 allow us to determine the
general structure of the amplitude of the isospin violating decay
$D^*_{s0}(2317)^+\to D_s^+\pi^0$. By the general structure we mean
the main contributions that form the decay $D^*_{s0}(2317)^+\to
D_s^+\pi^0$. Using tentative values for the parameters at our
disposal, we present at the end of Sec. III numerical estimates for
the decay width $\Gamma(D^*_{s0}(2317)^+\to D_s^+\pi^0)$. Brief
conclusions from the analysis performed are presented in Sec. IV.
Several cumbersome formulas are placed in the Appendix.


\section{\boldmath Unmixed amplitudes with $I=0$ and 1}

\subsection{\boldmath $I=0$ sector}

Consider a state with the open charm $C=+1$, strangeness $S=+1$,
isospin $I=0$, spin-parity $J^P=0^+$, and mass $m_r\simeq2317.8$ MeV
associated with closed $D^0K^+$, $D^+K^0$, and $D_s^+\eta$ channels
(the thresholds of these channels are 2358.517 MeV, 2367.111 MeV,
and 2516.212 MeV, respectively). We denote it as $D^*_{s0}(2317)^+$.
In the limit of isotopic invariance, such a state is stable.
However, $\pi^0-\eta$ mixing leads to the possibility of its decay
with violation of isotopic invariance to the $D_s^+\pi^0$ channel,
the threshold of which is 2103.327 MeV, as well as to the
possibility of $D_s^+\pi^0$ scattering via the $D^*_{s0}(2317)^+$
intermediate state [see diagrams (a) and (b) in Fig. 1].
\begin{figure}  [!ht] 
\begin{center}\includegraphics[width=7.75cm]{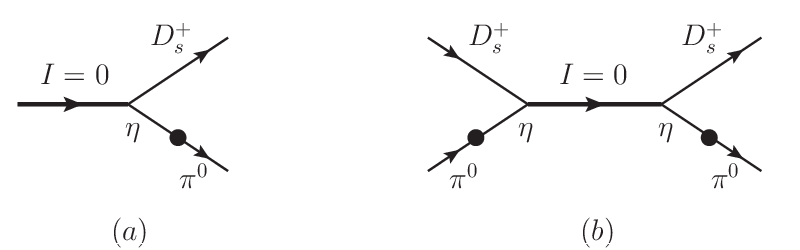}
\caption{\label{Fig1} (a) Decay of the $I=0$ state, $D^*_{s0}(2317
)^+$, into $D_s^+\pi^0$. (b) $D_s^+\pi^0$ scattering via the
$D^*_{s0}(2317)^+$ intermediate state resulting from the
$\pi^0-\eta$ mixing. Each black circle in this figure, as well as in
Figs. 2, 3, and 6, denotes the $\pi^0-\eta$ mixing amplitude
$\Pi_{\pi^0\eta}$}. \end{center}\end{figure}
\begin{figure}  [!ht] 
\begin{center}\includegraphics[width=12cm]{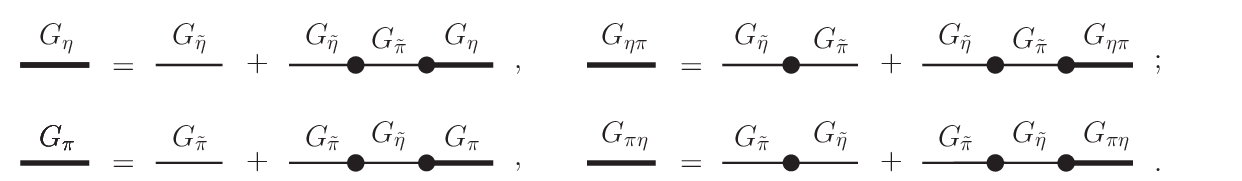}
\caption{\label{Fig2} Equations for propagators of the mixed $\eta$
and $\pi^0$ mesons and propagators of the $\pi^0 \leftrightarrow
\eta$ transitions; the argument $q^2$, on which the propagators
depend, is omitted in the figure (the notations are explained in
detail in the text).} \end{center}\end{figure}
The coupling of the isoscalar $D^*_{s0}(2317 )^+$ state with the
$D_s^+\pi^0$ system due to the $\pi^0-\eta $ mixing leads to a
corresponding contribution to its polarization operator. In order to
correctly calculate this contribution in the second order in the
amplitude of the $\pi^0-\eta$ mixing, $\Pi_{\pi^0 \eta}$, we
consider the system of equations graphically depicted in Fig. 2 for
the propagators of the mixed $\eta$ and $\pi^0$ mesons and the
propagators of the $\pi^0 \leftrightarrow \eta$ transitions denoted
as $G_\eta(q^2)$, $G_\pi(q^2)$ and $G_{\eta\pi}(q^2)=G_{\pi\eta}
(q^2)$, respectively. The propagators of unmixed $\tilde{\eta}$ and
$\tilde{\pi}^0$ mesons are $G_{\tilde{\eta}}(q^2)=1/ D_{\tilde{\eta}
}(q^2)=1/(m^2_{ \tilde{\eta}}-q^2)$ and $G_{\tilde{\pi} }(q^2)
=1/D_{\tilde{\pi}}(q^2)=1/(m^2_{\tilde{\pi}}-q^2)$. Solving the
equations in Fig. 2, we find
\begin{eqnarray}\label{Eq1}
G_\eta(q^2)=\frac{D_{\tilde{\pi}}(q^2)}{D_{\tilde{\eta}}(q^2)D_{
\tilde{\pi}}(q^2)-\Pi^2_{\pi^0\eta}},\ \, G_\eta(q^2)=\frac{D_{
\tilde{\eta}}(q^2)}{D_{\tilde{\eta}}(q^2)D_{\tilde{\pi}}(q^2)-
\Pi^2_{\pi^0\eta}},\ \, G_{\eta\pi}(q^2)=G_{\pi\eta} (q^2)=\frac{
\Pi_{\pi^0\eta}}{D_{\tilde{\eta}}(q^2)D_{\tilde{\pi}}(q^2)-\Pi^2_{
\pi^0\eta}}. \end{eqnarray} Further the easiest way to proceed is as
follows. Consider, for example, for $G_\eta(q^2)$ the chain of
equalities:
\begin{eqnarray}\label{Eq2}
G_\eta(q^2)=\frac{1}{D_{\tilde{\eta}}
(q^2)-\frac{\Pi^2_{\pi^0\eta}}{D_{\tilde{\pi}}(q^2)}}=\frac{1}{
D_\eta(q^2)-\frac{\Pi^2_{\pi^0\eta}}{D_{\tilde{\pi}}(q^2)}
+\frac{\Pi^2_{\pi^0\eta}}{D_{\tilde{\pi}}(m^2_\eta)}}\qquad\qquad
\qquad\qquad \nonumber \\
=\frac{1}{D_\eta(q^2)}+\frac{\Pi_{\pi^0\eta}}{D_\eta(q^2)} \left(
\frac{1}{D_\pi(q^2)}-\frac{1}{D_\pi(m^2_\eta)}\right)
\frac{\Pi_{\pi^0\eta}}{D_\eta(q^2)}=\frac{1}{D_\eta(q^2)}+\frac{
\Pi_{\pi^0\eta}}{m^2_\eta-m^2_\pi}\left(\frac{\Pi_{\pi^0\eta}}{
D_\eta(q^2)D_\pi(q^2)}\right).\end{eqnarray} Here in the second
equality the $\tilde{\eta}$ meson mass has been renormalized in the
second order of perturbation theory in $\Pi_{ \pi^0\eta}$ and the
notation $D_\eta(q^2)=m^2_\eta-q^2$ is introduced, where $m^2_\eta$
is the renormalized (physical) mass of $\eta$. Further, without
violating the accuracy of the approximation, we can use the physical
value for the $\pi$ meson mass everywhere, i.e., replace
$D_{\tilde{\pi}}(q^2)$ by $D_\pi(q^2)=m^2_\pi-q^2$. The third and
fourth equalities are differently written first terms of the
expansion of $G_\eta(q^2)$ in $\Pi^2_{\pi^0\eta}$ (which operate
within the second order of perturbation theory). Note that the
expression in parentheses in the last equality is the transition
propagator $G_{\eta\pi}(q^2)$ written in the first order in
$\Pi_{\pi^0\eta}$. The expressions for the propagator $G_\eta(q^2)$
and similar expressions for the propagator $G_\pi(q^2)$ obtained to
the second order in $\Pi_{\pi^0\eta}$ will be used below in
calculating the polarization operators.

Let us proceed to the definition of the polarization operator of the
$D^*_{s0}(2317)^+$ resonance. For short, we denote the channels
$D^0K^+$, $D^+K^0$, $D_s^+\eta$, and $D_s^+\pi^0$ by numbers 1, 2,
3, and 4, respectively. The $S$-wave amplitudes $T^{I}_{ij}(s)$ for
the reactions $i\to j$ corresponding to the contribution of the
$D^*_{s0}(2317)^+$ resonance with $I=0$ are written in the form
\begin{eqnarray}\label{Eq3}
T^{0}_{ij}(s)=\frac{g^{0}_ig^{0}_j}{D^{0}_r(s)}
=\frac{g^{0}_ig^{0}_j}{m^2_r-s+\mbox{Re}\,\Pi^{0}_r(m^2_r)-\Pi^{0
}_r(s)},\end{eqnarray} where $i$ and $j$ are the channel numbers,
$s$ is the invariant mass squared in channel $i$ and $j$,
$1/D^{0}_r(s)$ is the propagator of the $D^*_{s0}(2317)^+$
resonance, $g^{0}_i$ is its coupling constant with channel $i$, in
so doing $g^{0}_4=\epsilon g^{0}_3$, and $\epsilon=\Pi_{\pi^0\eta}/
(m^2_\eta-m^2_{\pi^0})$ is the parameter of the $\pi^0-\eta$ mixing.
Here we will be guided on the value of $\epsilon\simeq-0.014$
\cite{Fe00,Io01}. The polarization operator $\Pi^{0}_r(s)$ in
(\ref{Eq3}) is the sum of the self-energy parts of the resonance
$D^*_{s0}(2317)^+$ due to the $D^0K^+$, $D^+K^0$, $D_s^+\eta$, and
$D_s^+\pi^0 $-intermediate states, i.e.,
\begin{eqnarray}\label{Eq4}
\Pi^{0}_r(s)=\sum_{i=1,2,3}\frac{(g^{0}_i)^2}{16\pi}G_{ii}(s)+
\frac{(g^{0}_3)^2}{16\pi}\ddot{G}_{44}(s),\end{eqnarray} where the
functions $G_{ii}(s)$ for $i=1,2,3$ are defined as the dispersion
two-body loop integrals subtracted at the threshold of the $i$th
channel. Explicit expressions for them are written out in the
Appendix.
\begin{figure}  [!ht] 
\begin{center}\includegraphics[width=12cm]{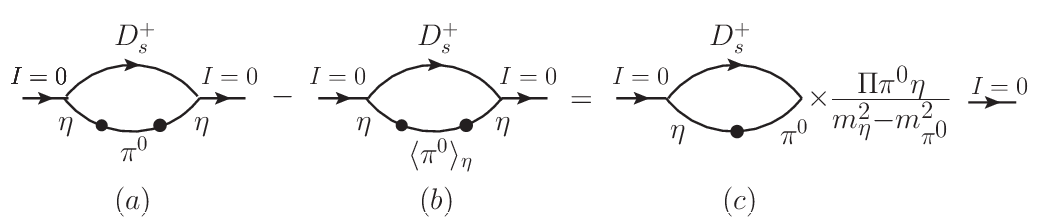}
\caption{\label{Fig3} According to Eq. (\ref{Eq2}) there are two
equivalent ways of calculating the function $\ddot{G}_{44}(s)$ using
the right-hand or left-hand side of the equality shown in the
figure. In diagram (b), $\langle \pi^0\rangle_\eta$ denotes that the
pion proragator $1/D_\pi(q^2)$ is taken at $q^2=m^2_\eta$.}
\end{center} \end{figure}
The two points in the notation of the function $\ddot{G}_{44}(s)$
indicate that we are dealing with the second-order contribution in
the amplitude $\Pi_{\pi^0\eta}$. According to the last two
equalities in Eq. (\ref{Eq2}), the function $\ddot{G}_{44}(s)$ can
be calculated using the diagrams shown in Fig. 3 in two equivalent
ways, i.e., using either the right-hand or the left-hand side of the
equality shown in the figure. We write the amplitude of the
right-hand side of this equality in the form
$\dot{G}_{34}(s)\frac{\Pi_{\pi^0\eta}}{m^2_\eta-m^2_{\pi^0}}$, where
$\dot{G}_{34}(s)$ denotes the first-order in the $\Pi_{\pi^0\eta}$
convergent loop diagram, diagram (c), shown in the figure. An
explicit expression for $\dot{G}_{34} (s)$ is given in the Appendix;
$\dot{G}_{34}(s)=\dot{G}_{43}(s)$. Thus, $\ddot{G}_{44}(s)$ can be
calculated by the rule $\ddot{G}_{
44}(s)=\dot{G}_{34}(s)\frac{\Pi_{\pi^0\eta}}{m^2_\eta- m^2_{\pi^0
}}$. Calculating $\ddot{G}_{44}(s)$ using diagrams (a) and (b),
which define the left-hand side of the equality in Fig. 3, turns out
to be more complicated. It is worth noting that at the $D^+_s\eta$
threshold, the imaginary parts of diagrams (a) and (b) in Fig. 3
have root threshold singularities, i.e., they contain contributions
$\sim1/\sqrt{s-(m_{D^+_s}+m_\eta)^2}$ that cancel out in the
difference of these diagrams.

The decay width of the resonance, $\Gamma_r(s)$, is determined by
the imaginary part of the polarization operator. In this case, in
the region of 2317 MeV, $\sqrt{s}\Gamma_r(s)=\mbox{Im}\,\Pi^{0
}_r(s)=\frac{(g^0_3)^2}{
16\pi}\mbox{Im}\,\ddot{G}_{44}(s)=\frac{(g^0_3)^2}{
16\pi}\frac{\Pi_{\pi^0\eta}}{m^2_\eta- m^2_{\pi^0}}
\mbox{Im}\,\dot{G}_{34}(s)$ and therefore, due to the mechanism of
the $\pi^0-\eta$ mixing, we have (see Fig. 3 and Appendix)
\begin{eqnarray} \label{Eq5} \sqrt{s}\Gamma_r(s)=\frac{
(g^0_3)^2}{16\pi}\left(\frac{\Pi_{\pi^0\eta}}{m^2_\eta-m^2_{
\pi^0}}\right)^2\rho_{D^+_s\pi^0}(s),\end{eqnarray} where
$\rho_{D^+_s\pi^0}(s)=\sqrt{[s-(m_{D^+_s}+m_{\pi^0})^2]
[s-(m_{D^+_s}-m_{\pi^0})^2]}/s$. Diagram (a) in Fig. 1 leads to
exactly the same expression for $\sqrt{s}\Gamma_r(s)$ (in full
agreement with the unitarity condition). Thus, we have before us a
kind of ``classical'' resonance, which could exist in the case of
the absence (or extreme smallness) of strong interactions in the
$D^0K^+$, $D^+K^0$, and $D_s^+\pi^0$ channels with isospin $I=1$.
Literally, such a situation seems unlikely. Consideration at this
stage of the $D^*_{s0}(2317)^+$ meson as an isolated state with
$I=0$ suggests that the amplitudes due to its contribution will be
used to construct a more realistic picture of interactions in the
coupled channels under consideration.


Let us now introduce some new notations that will be useful later.
We write the coupling constants $g^{0}_i$ as $g^{0}_i=g^{0
}_1\eta^{0}_i$, where $\eta^{0}_i=g^{0}_i/g^{0}_1$ are dimensionless
constants on the scale of $g^{0}_1$, and define the reduced
(dimensionless) propagator of the $D^*_{s0}(2317)^+$ resonance
$1/\tilde{D}^{0}_r (s)=1/[D^{0}_r(s)/(g^{0}_1)^2]$. In these
notations $T^{0 }_{ij}(s)=\eta^{0}_i\eta^{0}_j\tilde{D}^{0}_r(s) $.
Due to isotopic invariance, $\eta^{0}_1=\eta^{0}_2=1$ and, by
definition (see above), $\eta^{0}_4=\epsilon\eta^{0}_3$. The value
of $\eta^{0}_3$ can be found within the framework of a specific
model of the interaction of the isoscalar $D^*_{s0}(2317)^+$ state
with the channels $D^0K^+$, $D^+K^0$, and $D_s^+\eta$. For example,
in the $c\bar s$ model, the relative values of the coupling
constants $D^*_{s0}(2317)^+$ with the channels $D^0K^+$, $D^+K^0$,
and $D_s^+\eta$ are easily determined using the diagram in Fig. 4.
\begin{figure}  [!ht] 
\begin{center}\includegraphics[width=5cm]{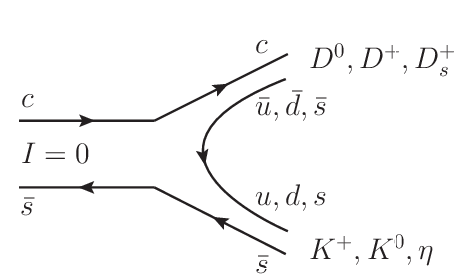}
\caption{\label{Fig4} Coupling of the $c\bar s$ state with channels
$D^0K^+$, $D^+K^0$, and $D_s^+\eta$.}
\end{center} \end{figure}
The usual quark counting rule, together with the representation of
the final $s\bar s$ pair in terms of physical states $\eta$ and
$\eta'$ with defined masses ($s\bar s=-\eta\cos\vartheta
+\eta'\sin\vartheta$, where $\vartheta=\theta_i-\theta_p$,
$\theta_i=35.3^\circ$ is the so-called ideal mixing angle and
$\theta_p=-11.3^\circ$ is the mixing angle in the nonet of light
pseudoscalar mesons \cite{PDG24}), leads to the relations
$\eta^{0}_1:\eta^{0}_2:\eta^{0}_3= 1:1:-\cos\vartheta=1:1:-0.687$.
If all $\eta^{0}_i$ are known, then an estimate of the overall
constant $(g^{0}_1)^2$ can be obtained using the available (so far
only theoretical) values of the $DK$ scattering length,
$a^{(0)}_{DK}$, in the channel with $I=0$
\cite{Gu09,Li13,Os15,Gu19,Hu22,Os23,To23}. In accordance with the
calculations \cite{Gu09,Li13,Os15,Gu19,Hu22, Os23,To23} we will be
guided by $a^{(0)}_{DK}\approx-1$ fm. Let us express $a^{(0)}_{DK}$
through the amplitude $T^{0}_{11}(s)+T^{0}_{ 12}(s)=2 T^{0}_{11}(s)$
at the $D^0K^+$ threshold. In our normalization we have
\footnote{Note that for the given mass of the bound state $m_r$, the
possible values of $a^{(0)}_{DK}<0$ (due to its contribution) are
limited from below by the value $-1.432$ fm, which is obtained if in
$T^{0}_{11}((m_{D^0}+ m_{K^+})^2)$ we let $(g^{0}_1)^2$ tend to
infinity.}
\begin{eqnarray}\label{Eq6} a^{(0)}_{DK}=\frac{2T^{0}_{11}(s)}
{8\pi\sqrt{s}}\left|_{\sqrt{s}=(m_{D^0}+ m_{K^+})}\right.\approx-1
\,\mbox{fm}. \end{eqnarray} From here we find $(g^{0}_1)^2/(16\pi)
\approx1.884$ GeV$^2$ and from Eq. (\ref{Eq5}) obtain
$\Gamma_r(m^2_r)\approx19.3$ keV. In estimating $(g^{0}_1)^2/(16
\pi)$, we neglected the negligible effect of the $\pi^0-\eta$
mixing. Thus, we have completely determined the resonant amplitudes
$T^{0}_{ij}(s)$.

\subsection{\boldmath $I=1$ sector}

Let us proceed to the construction of $S$-wave nonresonance
amplitudes of the reactions $i\to j$ with isospin $I=1$ in the
$s$-channel $T^{1}_{ij}(s)$. As a guide, we will use the Zachariazen
field theoretical model for the single-channel $S$-wave scattering
amplitude that was once thoroughly investigated in the works
\cite{Za61, GZ61,Th62,Za65}. The amplitude $T(s)$ in the Zachariasen
model exactly coincides with the result of the summing up of all
chains of $s$-channel loop diagrams in the theory with Lagrangian
$\mathcal{L}=-\lambda_0\varphi^4_a$, where $\lambda_0$ is the seed
coupling constant and $\varphi_a$ is the scalar field with mass
$m_a$ \cite{Za61,Th62,Za65}. In terms of renormalized quantities the
amplitude $T(s)=-\lambda/[1+\frac{\lambda}{16\pi }G(s)]$, where
$\lambda$ is the coupling constant and $G(s)$ is the dispersion loop
integral once subtracted (for definiteness) at $s=4m^2_a$. For
positive $\lambda$, this model produces a ``dynamic'' bound state
\cite{GZ61,Th62}. But for negative values of $\lambda$, the model
gives a good example of the nonresonance scattering amplitude and
phase \cite{GZ61,Th62}. We will use one of the simplest
generalizations of such an amplitude for the case of several coupled
channels with $I=1$, which looks like this:
\begin{eqnarray}\label{Eq7}
T^{1}_{ij}(s)=\frac{-\lambda\eta^{1}_i\eta^{1}_j}{D^{1}_{n}(s)}=
\frac{-\lambda\eta^{1}_i\eta^{1}_j}{1+\Pi^{1}_n(s)},\end{eqnarray}
where the index $n$ indicates that the corresponding quantities
describe the nonresonance case, and the polarization operator
$\Pi^{1}_n(s)$ has the form
\begin{eqnarray}
\label{Eq8}\Pi^{1}_n(s)=\frac{\lambda}{16\pi}\left[\sum_{i=1,2,4}(\eta^{1}_i)^2G_{ii}
(s)+(\eta^{1}_4)^2\ddot{G}_{33}(s)\right]. \end{eqnarray} Here
$G_{11}(s)$, $G_{22}(s)$, and $G_{44}(s)$ are the dispersion loop
integrals subtracted at the corresponding thresholds and the
function $\ddot{G}_{33}(s)=-\ddot{G}_{44}(s)=-\dot{G}_{34}
(s)\frac{\Pi_{\pi^0 \eta}}{m^2_\eta-m^2_{\pi^0}}$ (see Appendix);
$\lambda$ is the coupling constant; $\eta^{1}_1=-\eta^{1}_2=1/
\sqrt{2}$, $\eta^{1}_4 =1$; in (\ref{Eq7}),
$\eta^{1}_3=-\epsilon\eta^{1}_4$. Let us explain the details of such
a representation of $T^{1}_{ij}(s)$. In the multichannel case, the
seed constant $\lambda_0$ is replaced by a symmetric matrix
$\lambda_{0ij}$ composed of seed amplitudes of the transitions $i\to
j$. Due to isotopic invariance, this matrix has the form
\begin{eqnarray}
\label{Eq9}\lambda_{0ij}=\left(\begin{array}{ccc}
\lambda_{011} & -\lambda_{011} & \lambda_{041} \\
-\lambda_{011} & \lambda_{011} & -\lambda_{041} \\
\lambda_{041} & -\lambda_{041} & \lambda_{044} \\
\end{array}\right).\end{eqnarray}
It contains three independent constants $\lambda_{011}$,
$\lambda_{041}$, and $\lambda_{044}$ corresponding to the
transitions $D^0K^+\to D^0K^+$, $D^+_s\pi^0\to D^0K^+$, and
$D^+_s\pi^0\to D^+_s\pi^0$, respectively. Note that in the sector
with $C=S=+1$ and $I=1$ all intermediate states must be at least
four-quark ones. We assume that the seed interaction
\begin{figure}  [!ht] 
\begin{center}\includegraphics[width=5cm]{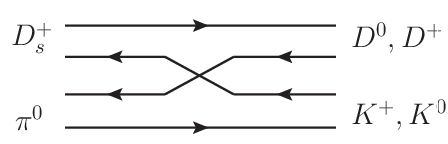}
\caption{\label{Fig5} Example diagram of the quark rearrangement
during scattering.}\end{center} \end{figure}
between the continuous spectrum states $D^0K^+$, $D^+K^0$, and
$D_s^+\pi^0$ is carried out due to the rearrangement of valence
quarks in colliding particles, see Fig. 5. In this case, the matrix
$\lambda_{0ij}$ has a very simple form
\begin{eqnarray}
\label{Eq9a}\lambda_{0ij}=\left(\begin{array}{ccc}
0 & 0 & \lambda_0 \\
0 & 0 & -\lambda_0 \\
\lambda_0 & -\lambda_0 & 0 \\
\end{array}\right)\end{eqnarray}
since the rearrangement mechanism allows only the transitions
$D_s^+\pi^0\leftrightarrow D^0K^+$ and $D_s^+\pi^0\leftrightarrow
D^+K^0$ the amplitudes of which due to isotopic invariance differ in
sign. Nevertheless, expressions for the amplitudes $T^{1}_{ij}(s)$
in the original unsubtracted form are far from trivial. In matrix
notation, the result of summing up all chains of the $s$ channel
loop diagrams in the model with a seed interaction from Eq.
(\ref{Eq9a}) has the form $\hat{T}^{1}(s)=-[ \hat{I}
+\hat{\lambda}_0 \hat{\Delta}(s)]^{-1}\hat{\lambda}_0$, where
$\hat{\Delta}(s)=B \hat{I}+\hat{G}(s)$, $B$ is an infinite constant,
and $\hat{G}$ is the diagonal matrix of the dispersion loop
integrals subtracted at the corresponding thresholds \cite{Th62}.
Renormalization in the amplitudes $T^{1}_{ij}(s)$ is carried out
using the relation $B(1-\frac{1}{2B^2\lambda^2_0})=1/\lambda$, where
$B\to\infty$, $\lambda_0\to0$, but so that $\lambda$ is a finite
value. After taking into account the $\pi^0-\eta$ mixing [similarly
to how it was done in the case of the amplitudes $T^{0}_{ij}(s)$] we
obtain Eqs. (\ref{Eq7}) and (\ref{Eq8}). We also introduce the
reduced propagator $1/\tilde{D}^1_n(s)=1/[D^1_n(s)/(-\lambda)]$. In
these terms $T^{1}_{ij}(s)=\eta^{1}_i\eta^{1}_j/\tilde{D}^1_n(s)$.

Thus, we have constructed the resonance amplitudes $T^{0}_{ij}(s)$
in the sector with isospin $I=0$ and the nonresonance ones
$T^{1}_{ij}(s)$ in the sector with isospin $I=1$, which do not yet
know anything about each other. Now we are ready to move on to
considering the mixing of these amplitudes.

\section{\boldmath Mixed $D^*_{s0}(2317)^+$ state}

The amplitudes $T_{ij}$ that would take into account the mixing of
the isoscalar resonance $D^*_{s0}(2317)^+$ with the nonresonance
amplitudes with isospin $I=1$ can be constructed in the same way as
was done when considering  phenomena of the $a_0(980)-f_0(980)$
mixing \cite{ADS79}. Graphically, the scheme for accounting for
mixing in the form of equations for the reduced propagators dressed
by mixing $\tilde{ \mathcal{G}}_0(s)$ and $\tilde{\mathcal{G}}_1
(s)$ in the channels with $I=0$ and 1, respectively, and for the
propagators of the transition between these channels
$\tilde{\mathcal{G} }_{01}(s)$ and $\tilde{\mathcal{G}}_{10}(s)$
looks completely similar to the scheme for accounting for mixing
shown in Fig. 2. From the corresponding equations we find that the
propagators dressed by mixing have the form
\begin{eqnarray}
\label{Eq10} \tilde{\mathcal{G}}_0(s)=\frac{\tilde{D}^1_n(s)}{
\tilde{D}^0_r(s)\tilde{D}^1_n(s)-\tilde{\Pi}^2_{01}(s)},\ \
\tilde{\mathcal{G}}_1(s)=\frac{\tilde{D}^0_r(s)}{
\tilde{D}^0_r(s)\tilde{D}^1_n(s)-\tilde{\Pi}^2_{01}(s)},\ \
\tilde{\mathcal{G}}_{01}(s)=\tilde{\mathcal{G}}_{10}(s)=
\frac{\tilde{\Pi}_{01}(s)}{\tilde{D}^0_r(s)\tilde{D}^1_n(s)-
\tilde{\Pi}^2_{01}(s)}\end{eqnarray} Here the functions
$\tilde{\Pi}_{01}(s)$ and $\tilde{\Pi}_{10}(s)$ are polarization
operators that are nondiagonal in isotopic spin, responsible for
mixing channels with $I=0$ and 1;
$\tilde{\Pi}_{01}(s)=\tilde{\Pi}_{10}(s)$. The polarization operator
$\tilde{\Pi}_{01}(s)$ has the form
\begin{eqnarray}
\label{Eq11}\tilde{\Pi}_{01}(s)=\frac{1}{16\pi}\left[\eta^0_3
\dot{G}_{34}(s) \eta^1_4+\eta^0_1\hat{G}_{11}(s)\eta^1_1+\eta^0_1
\hat{G}_{22}(s)\eta^1_2\right] \nonumber \\
=\frac{1}{16\pi}\left[-\cos\vartheta\,\dot{G}_{34}(s)+\frac{1}{
\sqrt{2}}\hat{G}_{11}(s)-\frac{1}{\sqrt{2}}\hat{G}_{22}(s)\right].
\end{eqnarray}
It is formed by two main sources of isotopic invariance violation,
which are the $\pi^0-\eta$ mixing and the presence of the mass
differences between charged and neutral $D$ mesons and kaons. Thanks
to  the latter, the contributions of the $D^0K^+$ and $D^+K^0$
intermediate states, entering to $\tilde{\Pi}_{01}(s)$ with opposite
signs, do not completely compensate each other. The intermediate
states in $\tilde{\Pi}_{01}(s)$ do not have a definite isotopic spin
and therefore contribute to mixing. The contributions to
$\tilde{\Pi}_{01}(s)$ are described by convergent expressions. The
function $\dot{G}_{34}(s)$ has already been encountered in Sec. II
(see, in particular, Fig. 3). In $\tilde{\Pi}_{01}(s)$, it is
responsible for the contribution caused by the $\pi^0-\eta$ mixing.
The difference between the $D^0K^+$ and $D^+K^0 $ loops
$\hat{G}_{11}(s)-\hat{G}_{2 2}(s)$ [see the second and third terms
in Eq. (\ref{Eq11})] also does not contain divergence. In so doing,
$\hat{G}_{11}(s)-\hat{G}_{2 2}(s)$ differs from the difference of
the integrals $G_{11}(s)$ and $G_{22}(s)$, subtracted at the
corresponding thresholds, by a constant depending on the ratios of
the masses of the particles in the loops (see Appendix).

Using the propagators $\tilde{\mathcal{G}}(s)$ from Eq.
(\ref{Eq10}), it is easy to construct the amplitudes $T_{ij}$. The
result will be exactly the same as when opening the expression
\begin{eqnarray}
\label{Eq12}T_{ij}=\eta^I_i \left(\begin{array}{cc}
\tilde{D}^0_r(s) & -\tilde{\Pi}_{01}(s) \\
-\tilde{\Pi}_{10}(s) & \tilde{D}^1_n(s) \\
\end{array}\right)^{-1}_{I,I'}\eta^{I'}_j.\end{eqnarray}
Thus we get
\begin{eqnarray}
\label{Eq13}T_{ij}=\frac{\eta^0_i\tilde{D}^1_n(s)\eta^0_j+\eta^0_i
\tilde{\Pi}_{01}(s)\eta^1_j+\eta^1_i\tilde{\Pi}_{10}(s)\eta^0_j+
\eta^1_i\tilde{D}^0_r(s)\eta^1_j}{\tilde{D}^0_r(s)\tilde{D}^1_n(s)-
\tilde{\Pi}^2_{01}(s)}.\end{eqnarray}

Let us consider in detail the amplitude $T_{44}(s)\equiv
T_{D^+_s\pi^0\to D^+_s\pi^0}(s)$ corresponding to the only opened
hadronic decay channel $D^+_s\pi^0$  in the region of
$D^*_{s0}(2317)^+$ resonance, and represent it as a sum of
{\it``background + resonance''}:
\begin{eqnarray}
\label{Eq14}T_{44}=\frac{\eta^0_4\tilde{D}^1_n(s)\eta^0_4+\eta^0_4
\tilde{\Pi}_{01}(s)\eta^1_4+\eta^1_4\tilde{\Pi}_{10}(s)\eta^0_4+\eta^1_4
\tilde{D}^0_r(s)\eta^1_4}{\tilde{D}^0_r(s)\tilde{D}^1_n(s)-
\tilde{\Pi}^2_{01}(s)}\qquad\qquad\nonumber \\
=\frac{\eta^1_4\eta^1_4}{\tilde{D}^1_n(s)}+\frac{\left(\eta^0_4+
\frac{\tilde{\Pi}_{01}(s)\eta^1_4}{\tilde{D}^1_n(s)}\right)^2}{\tilde{D}^0_r(s)-
\frac{\tilde{\Pi}^2_{01}(s)}{\tilde{D}^1_n(s)}}=\frac{16\pi}{\rho_{D^+_s\pi^0}(s)}
\left\{\frac{e^{2i\delta^1_0(s)}-1}{2i}+e^{2i\delta^1_0(s)}
T_{\scriptsize\mbox{res}}(s)\right\}.\end{eqnarray} The first term
in the second equality in Eq.(\ref{Eq14}) is the nonresonance
amplitude $T^1_{44}(s)=\eta^1_4\eta^1_4/\tilde{D}^1_n(s)=1/
\tilde{D}^1_n(s)= \frac{16\pi}{\rho_{D^+_s\pi^0}(s)}[-\mbox{Im}
\,\tilde{D}^1_n(s)/\tilde{D}^1_n(s)]= \frac{16\pi}{\rho_{D^+_s\pi^0}
(s)}e^{i\delta^1_0(s)}\sin\delta^1_0(s)$ or the background one, and
$\delta^1_0(s)$ is the background phase. The expression
$\eta^0_4+\frac{\tilde{\Pi}_{01}(s) \eta^1_4}{\tilde{D}^1_n(s)}$,
standing in brackets in the numerator of the second term in the
second equality in Eq. (\ref{Eq14}), is the amplitude of the decay
of the isoscalar resonance $D^*_{s0}(2317)^+$ (dressed by the
background) into $D^+_s\pi^0$ caused by the isospin breaking. We
denote it as $V_{\scriptsize\mbox{res}}(s)=\eta^0_4+\frac{
\tilde{\Pi}_{01}(s)\eta^1_4}{\tilde{D}^1_n(s)}$. Here the first term
$\eta^0_4=\epsilon\eta^0_3$ is due to the contribution of the first
diagram in Fig. 6 and the second one is due to the contributions of
the subsequent three diagrams in this figure in accordance with Eq.
(\ref{Eq11})
\begin{figure}  [!ht] 
\begin{center}\includegraphics[width=10.5cm]{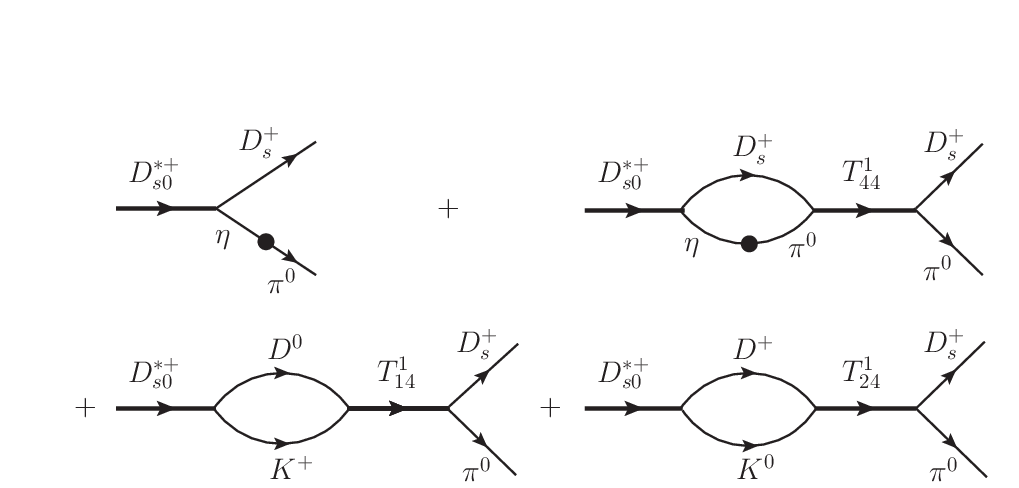}
\caption{\label{Fig6} Diagrams contributing to the decay amplitude
$D^*_{s0}(2317)^+\to D^+_s\pi^0$.}\end{center}
\end{figure}
for the polarization operator $\tilde{\Pi}_{01}(s)$ multiplied by
the amplitude $\frac{\eta^1_4}{\tilde{D}^1_n(s)}$ of the nonresonant
final state interaction in the channel with $I=1$. Using Eq.
(\ref{Eq7}), the relation $\mbox{Im}\,\dot{G}_{34}(s)= \epsilon
\rho_{D^+_s\pi^0} (s)$ (see Appendix), the above values of
$\eta^I_i$, and the relation of $1/\tilde{D}^1_n(s)$ with the phase
$\delta^1_0(s)$, we obtain the following expression for
$V_{\scriptsize\mbox{res}}(s)$:
\begin{eqnarray}
\label{Eq15}V_{\scriptsize\mbox{res}}(s)=e^{i\delta^1_0(s)}\left\{
-\epsilon\cos\vartheta\cos\delta^1_0(s)+\left[-\cos\vartheta\,
\mbox{Re}\,\dot{G}_{34}(s)+\frac{1}{\sqrt{2}}(\hat{G}_{11}(s)-\hat{G}_{2
2}(s))\right]\frac{\sin\delta^1_0(s)}{\rho_{D^+_s\pi^0}(s)}
\right\}=e^{i\delta^1_0(s)}v_{\scriptsize\mbox{res}}(s).\end{eqnarray}
As can be seen, the phase of the decay amplitude
$V_{\scriptsize\mbox{res}}(s)$ is determined by the phase of the
nonresonance $D_s^+\pi^0$ scattering amplitude in accordance with
the Watson theorem on the final state interaction \cite{Wa52} (or
with the unitarity  requirement). In the normalization we use, the
$D^*_{s0}(2317)^+\to D^+_s\pi^0$ decay width and the amplitude
$V_{\scriptsize \mbox{res}}(s)$ are connected by the relation
\begin{eqnarray} \label{Eq16}
\frac{\sqrt{s}\Gamma{\scriptsize\mbox{res}}(s)}{(g^0_1)^2}=\frac{|
V_{\scriptsize\mbox{res}}(s)|^2}{16\pi}\rho_{D^+_s\pi^0}(s).
\end{eqnarray}
It is easy to show that the imaginary part of the expression
$\tilde{D}^0_r(s)-\frac{\tilde{\Pi}^2_{01}(s)}{\tilde{D}^1_n(s)}$
[this is the denominator of the second term in the second equality
in Eq. (\ref{Eq14})] is exactly equal to $\frac{|V_{\scriptsize
\mbox{res}}(s)|^2}{16\pi}\rho_{ D^+_s\pi^0}(s)$ and, therefore,
determines the width of the resonance peak in the $D_s^+\pi^0$
channel. The expression $1/[\tilde{D}^0_r( s)-\frac{\tilde{\Pi}
^2_{01}(s)}{\tilde{D}^1_n(s)}]$ can naturally be called the reduced
propagator of the $D^*_{s0}(2317)^+$ resonance modified by the
background. The ratio $\frac{ \tilde{\Pi}^2_{01}(s)}{
\tilde{D}^1_n(s)}\equiv\tilde{\Pi}_{01}(s)\frac{1}{\tilde{D}^1_n(
s)}\tilde{\Pi}_{10}(s)$ included in it is a polarization operator of
the isoscalar state taking into account the isovector interaction
between particles in the loops. Its real part at
$\sqrt{s}=M_{\scriptsize\mbox{res}}$, where $M_{\scriptsize
\mbox{res}}$ is the  mass of the resonance dressed by the
background, together with the contribution $\frac{(\eta^{0}_3)^2}
{16\pi} \mbox{Re}\,\ddot{G}_{44}(M^2_{\scriptsize\mbox{res}})$ [see
Eqs. (\ref{Eq3}) and (\ref{Eq4})], determines the magnitude of the
mass shift due to isospin breaking (isotopic mass shift). Finally,
taking into account Eq. (\ref{Eq3}), we write out an expression for
the resonance amplitude $T_{\scriptsize\mbox{res}}(s)$ in Eq.
(\ref{Eq14})
\begin{eqnarray} \label{Eq17}
T_{\scriptsize\mbox{res}}(s)=\frac{\sqrt{s}\Gamma_{\scriptsize\mbox{res}}(s)}
{M^2_{\scriptsize\mbox{res}}-s+\mbox{Re}\,\Pi_{\scriptsize\mbox{res}}(M^2_{
\scriptsize\mbox{res}})-\Pi_{\scriptsize\mbox{res}}(s)}=\frac{\sqrt{s}
\Gamma_{\scriptsize\mbox{res}}(s)}{D_{\scriptsize\mbox{res}}(s)},
\end{eqnarray} where
\begin{eqnarray} \label{Eq18}
\Pi_{\scriptsize\mbox{res}}(s)=\Pi^{0}_r(s)+g^0_1\frac{
\tilde{\Pi}_{01}(s)\tilde{\Pi}_{10}(s)}{\tilde{D}^1_n(s)}g^0_1.
\end{eqnarray}

If $D^*_{s0}(2317)^+$ is produced by a source with isospin $I=0$,
then the amplitude of its production, propagation, and decay into
$D_s^+\pi^0$ can be written as
\begin{eqnarray} \label{Eq19}
T_{\scriptsize\mbox{prod}}(s)=\Lambda(s)\frac{e^{i\delta^1_0(s)}
\xi\sqrt{\sqrt{s}\Gamma_{\scriptsize\mbox{res}}(s)}}{M^2_{
\scriptsize\mbox{res}}-s+\mbox{Re}\,\Pi_{\scriptsize\mbox{res}}
(M^2_{\scriptsize\mbox{res}})-\Pi_{\scriptsize\mbox{res}}(s)},
\end{eqnarray}
where $\Lambda(s)$ is the source amplitude, $e^{i\delta^1_0(s)}$ is
the phase of the nonresonance $D_s^+\pi^0$ interaction in the final
state, and $\xi=v_{ \scriptsize\mbox{res}}(s)/
|v_{\scriptsize\mbox{res}} (s)|$ is the sign of the amplitude
$v_{\scriptsize\mbox{res}}(s)$, see (\ref{Eq15}). Formula
(\ref{Eq19}) follows from Eq. (\ref{Eq13}) when taking into account
in the numerator of the latter the first two terms corresponding to
the source with $I=0$.

Let us return to Eq. (\ref{Eq15}) for the amplitude $V_{\scriptsize
\mbox{res}}(s)$ and give the numerical values of the terms forming
it at $\sqrt{s}=M{\scriptsize\mbox{res}}\simeq2317.8$ MeV [the
values of $V_{\scriptsize \mbox{res}}(s)$ and $\delta^1_0(s)$ at
this point we denote as $\bar{V}_{\scriptsize \mbox{res}}$ and
$\bar{\delta}^1_0$, respectively]:
\begin{eqnarray} \label{Eq20}
\bar{V}_{\scriptsize\mbox{res}}=e^{i\bar{\delta}^1_0}\left\{0.009618
\cos\bar{\delta}^1_0+[0.001888+0.0101372]\frac{\sin\bar{\delta}^1_0
}{0.257048}\right\}=e^{i\bar{\delta}^1_0}\left(0.009618\cos
\bar{\delta}^1_0+0.0467801\sin\bar{\delta}^1_0\right).\end{eqnarray}
The value of $\bar{\delta}^1_0$ is unknown. But we can trace how the
decay width of $D^*_{s0}(2317)^+\to D_s^+\pi^0$ changes depending on
$\bar{\delta}^1_0$. The corresponding picture is shown in Fig. 7.
\begin{figure}  [!ht] 
\begin{center}\includegraphics[width=5.5cm]{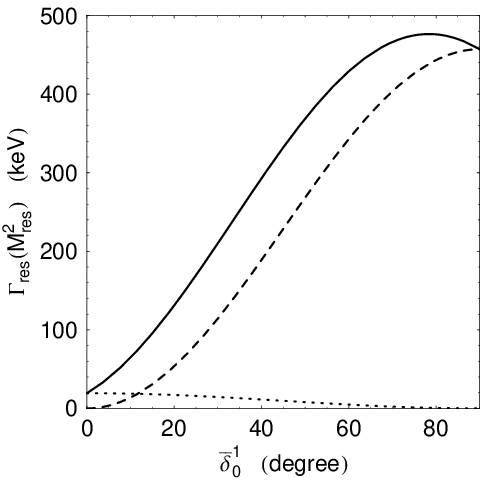}
\caption{\label{Fig7} Solid curve shows the decay width of
$D^*_{s0}(2317)^+\to D^+_s\pi^0$ at $\sqrt{s}=M_{\scriptsize
\mbox{res}}\simeq2317.8$ MeV as a function of $\bar{\delta}^1_0$,
constructed using Eqs. (\ref{Eq20}) and (\ref{Eq16}) at
$(g^0_1)^2/(16\pi)=1.884$ GeV$^2$. The dotted and dashed curves show
the contributions due to the $\pi^0-\eta$ mixing (proportional to
$\cos\bar{\delta}^1_0$) and meson loops (proportional to
$\sin\bar{\delta}^1_0$), respectively; see Eq.
(\ref{Eq15}).}\end{center}\end{figure}
Although large values of $\bar{\delta}^1_0$ are almost improbable,
Eq. (\ref{Eq20}) formally allows us to set an upper limit for
$\Gamma_{\scriptsize \mbox{res}}(M^2_{\scriptsize \mbox{res}})$. It
is $476.5$ keV at $\bar{\delta}^1_0=78.4^\circ$, see Fig. 7. The
values of the background phase $\bar{\delta}^1_0\approx(15-
20)^\circ$ can be considered quite reasonable. They lead to
$\Gamma_{ \scriptsize \mbox{res}}(M^2_{\scriptsize\mbox{res}
})\approx(95-130)$ keV (see Fig. 7). These values agree very well
with the calculations of $\Gamma_{ \scriptsize
\mbox{res}}(M^2_{\scriptsize\mbox{res}})$ based on the coupled
channel unitarized chiral perturbation theory and lattice QCD
\cite{Gu08,Li13,Cl14,Gu18,Fu22,Gu24}. Note that the isotopic mass
shift of the $D^*_{s0} (2317)^+$ resonance at $\bar{\delta}^1_0
=15^\circ$ and $20^\circ$ amounts to $-56$ keV and $-67$ keV,
respectively.

\section{Conclusion}

From the above analysis we draw the following conclusions.

1. The effect of isotopic invariance violation for the $D^*_{s0}
(2317)^+$ resonance is in many ways similar to the well-known
threshold phenomenon of the $a_0(980)^0$ and $f_0(980)$ resonance
mixing \cite{ADS79}.

2. The phenomenological approach we used allowed us to quite easily
determine the general structure of the amplitude of the isospin
violating decay of $D^*_{s0}(2317)^+\to D_s^+\pi^0$ and the
amplitude of the $S$-wave scattering process $D_s^+\pi^0\to
D_s^+\pi^0$ in the $D^*_{s0}(2317)^+$ resonance region [see
(\ref{Eq14}), (\ref{Eq15}), and Fig. 6]. Our numerical estimates for
the $D^*_{s0} (2317)^+\to D_s^+\pi^0$ decay width do not contradict
those available in the literature.

3. It is important that the constructed model expressions for the
complex of mixed amplitudes with $I=0$ and 1 entirely satisfy the
requirements of unitarity.

4. Our approach is fully applicable to the description of the mixing
of the $D^*_{s0} (2317)^+$ with the supposed resonance isovector
state $T_{c\bar s}(2327)^+$ \cite{Ma24,Aa25}. But here it is
necessary to note that the two-humped spectrum of $\pi^+\pi^-$ in
the decay of $D_{s1}(2460)^+\to D^+_s\pi^+\pi^-$, which gave a hint
about the existence of states $T_{c\bar s}(2327)$ with isospin $I=1$
\cite{Aa25}, can be explained without their introduction
\cite{Os25}.

5. It would be interesting to extend such a phenomenological
analysis to the isospin violating decay of $D_{s1}(2460)^+\to
D^{*+}_s \pi^0$. The presence of spin in particles in this case
introduces certain complications into the calculations with a purely
relativistic approach.


\vspace*{0.3cm}

\begin{center} {\bf ACKNOWLEDGMENTS} \end{center}

The work was carried out within the framework of the state contract
of the Sobolev Institute of Mathematics, Project No. FWNF-2022-0021.


\vspace*{0.3cm}

\begin{center} {\bf APPENDIX:\, THE FUNCTIONS {\boldmath
$G_{ii}(s)$, $\dot{G}_{34}(s)$, AND $\hat{G}_{11}(s)-\hat{G}_{22}(s)
$}}
\end{center}

\setcounter{equation}{0}
\renewcommand{\theequation}{A\arabic{equation}}


\setcounter{equation}{0}
\renewcommand{\theequation}{A\arabic{equation}}

The function $G_{ii}(s)$ is uniquely determined by the masses of
particles $a$ and $b$ in the $i$th channel. Let $m_a>m_b$. Then for
$i=1$ \,$m_a=m_{D^0}$ and $m_b=m_{K^+}$ etc. Let
$G_{ii}(s)=I_{ab}(s)$. The dispersion loop integral $I_{ab}(s)$ is
given by: for $s>m_{ab}^{(+)\,2}$
\begin{eqnarray}\label{A1} I_{ab}(s)=\frac{s-m_{ab}^{(+)\,2}}{\pi}
\int\limits^\infty_{m_{ab}^{(+)\,2}}\frac{\rho_{ab}(s')\,ds'}{
\,(s'-m_{ab}^{(+)\,2})(s'-s-i\varepsilon)}=L_{ab}(s)+\rho_{ab}(s)
\left(i-\frac{1}{\pi}\,\ln\frac{\sqrt{s-m_{ab}^{(-)
\,2}}+\sqrt{s-m_{ab}^{(+)\,2}}}{\sqrt{s-m_{ab}^{(-)\,2}}-\sqrt{s
-m_{ab}^{(+)\,2}}}\right),\
\end{eqnarray} where $m_{ab}^{(\pm)}=m_a\pm m_b$,
$\rho_{ab}(s)= \sqrt{s-m_{ab}^{(+)\,2}}
\,\sqrt{s-m_{ab}^{(-)\,2}}\,/s$,\,  and \,
\begin{eqnarray}\label{A2}L_{ab}(s)=\frac{1}{\pi}\left(\frac{s-m_{
ab}^{(+)\,2}}{s}\right) \frac{m_{ab}^{(-)}}{m_{ab}^{(+)}}\ln
\frac{m_a}{m_b};\end{eqnarray} for
$m_{ab}^{(-)\,2}<s<m_{ab}^{(+)\,2}$
$\rho_{ab}(s)$\,=\,$\sqrt{m_{ab}^{(+)\,2}-s}
\,\sqrt{s-m_{ab}^{(-)\,2}}\,/s$ and
\begin{eqnarray}\label{A3}
I_{ab}(s)=L_{ab}(s)-\rho_{ab}(s)\left(1-\frac{2}{\pi}
\arctan\frac{\sqrt{ m_{ab}^{(+)\,2}-s}}{\sqrt{s-m_{ab}^{
(-)\,2}}}\right); \end{eqnarray} for $s<m_{ab}^{(-)\,2}$
$\rho_{ab}(s)$\,=\,$\sqrt{m_{ab}^{(+)\,2}-s}\,\sqrt{
m_{ab}^{(-)\,2}-s}\,/s$ and
\begin{eqnarray}\label{A4} I_{ab}(s)=L_{ab}(s)+\frac{\rho_{ab}(s)
}{\pi}\,\ln\frac{ \sqrt{m_{ab}^{(+)\,2}-s}+\sqrt{m_{ab}^{(-)\,2
}-s}}{\sqrt{m_{ab}^{(+)\,2}-s}-\sqrt{m_{ab}^{(-)\,2}-s}}.
\end{eqnarray}
\begin{eqnarray}\label{A5} \dot{G}_{34}(s)=\frac{\Pi_{\pi^0\eta}}
{m^2_\eta-m^2_\pi}\left[G_{44}(s)-G_{33}(s)+\frac{1}{\pi}\left(\ln
\frac{m_\eta}{m_{\pi^0}}+\frac{m_{D^+_s}-m_{\eta}}{ m_{D^+_s}+m_{
\eta}}\ln\frac{m_{D^+_s}}{m_{\eta}}-\frac{m_{D^+_s}- m_{\pi^0}}{m_{
D^+_s}+m_{\pi^0}}\ln\frac{m_{D^+_s}}{m_{\pi^0}}\right)\right].
\end{eqnarray}
\begin{eqnarray}\label{A6}\hat{G}_{11}(s)-\hat{G}_{22}(s)=
G_{11}(s)-G_{22}(s)+\frac{1}{\pi}\left(\ln \frac{m_{D^+}m_{K^0}}
{m_{D^0}m_{K^+}}+\frac{m_{D^+}-m_{K^0}}{m_{D^+}+m_{K^0}}\ln\frac{
m_{D^+}}{m_{K^0}}-\frac{m_{D^0}-m_{K^+}}{m_{D^0}+m_{K^+}}\ln\frac{
m_{D^0}}{m_{K^+}}\right).
\end{eqnarray}

Let us explain the origin of the sign in the relation
$\ddot{G}_{33}(s) =-\ddot{G}_{44}(s)=-\dot{G}_{34}
(s)\frac{\Pi_{\pi^0 \eta}}{m^2_\eta-m^2_{\pi^0}}$ [see the text
below Eq. (\ref{Eq8})]. The function $\ddot{G}_{33}(s)$ can be
calculated in two ways, using either the right or left sides of the
equality shown in Fig. 3, after replacing $\eta$ with $\pi^0$,
$\pi^0$ with $\eta$, and $I=0$ with $I=1$. The function
$\dot{G}_{34}(s)$ is symmetric with respect to such a replacement
[see (\ref{A5})]. Therefore, the right-hand side of this equality as
a whole changes sign. The change in sign of its left-hand side after
the indicated replacement can be verified by direct calculation.


\end{document}